\documentstyle[12pt]{article}
\begin{document}
\textwidth 160mm
\textheight 240mm
\topmargin -20mm
\oddsidemargin 0pt
\evensidemargin 0pt
\newcommand{\beq}{\begin{equation}}
\newcommand{\eeq}{\end{equation}}
\begin{titlepage}
\begin{center}

{\bf Multigluon Tree Amplitudes and  Self-Duality Equation}

\vspace{1.5cm}

{\bf  K.G.Selivanov }

\vspace{1.0cm}

{ITEP,B.Cheremushkinskaya 25,Moscow,Russia}

\vspace{1.9cm}

{ITEP-21/96}

\vspace{1.0cm}

\end{center}


\begin{abstract}
A generating function for a class of multigluon amplitudes is constructed as a particular
solution of the self-duality equation.
\end{abstract}
\end{titlepage}


\newpage

1. Recently, there has been developed an interesting activity based
on the fact that a  soliton solution in a generic massive field
theory happens to be a generating function for three multiparticle
amplitudes with most of the particles at threshold  (see reviews \cite{volrub}). The
corresponding technique essentially improves the perturbative
expansion for the threshold multiparticle processes \cite{smith}, \cite{arg} and sometimes allows to
take into account all loop corrections \cite{bez}.  Another interesting
result having come from that approach is very nontrivial
nullifications among the tree threshold amplitudes \cite{voloshin}. The
nullification was  shown to survive in some theories at the quantum
level \cite{D}. In a particular model it was explaned by a hidden
conservation law  \cite{E} and it was given a general
interpretation in terms of the algebro-geometric approach to
solitonic equations in \cite{GS}.

On the other hand  side, there has been developed an involved
technology for efficient calculating the multigluon amplitudes in Yang-Mills (YM) theory,
basically, at tree level (reviewed in \cite{mapa}) and at one-loop level (reviewed, e.g.,
in \cite{dixon}). One of the main ingredients is the  spinor helicity formalism \cite{G}.
Interestingly, the multigluon amplitudes have a counterpart of the
nullification mentioned above. Namely, the tree amplitudes with all
and with all but one external gluons having identical helicities are
zero \cite{mapa}.

This note is aimed to bring these two seemingly different activities
together by demonstrating that a generating function for the
multigluon amplitudes can be obtained by solving the self-duality
(SD) equations.

I consider here only the SU(2) case and use the simple 't Hooft
anzatz to deal with the SD equation. The SU(N) case will be
considered separately with use of the twistors for constructing the
appropriate solution, which naturally lead to the spinor helicity
nitti gritti \cite{G}.

2. The starting point is the following version of the
Lehmann-Symanzik-Zimmerman formula (see also \cite{brown}):

\begin{eqnarray}
\left( \frac{\delta S_{tree}(\{a,a^*\})}{\delta a^j_{\lambda}(p)}
\right)_c = i\int d^4 x \frac{1}{2\omega_p} e^{-ipx}
\varepsilon^{\lambda}_{\mu} \cdot  \nonumber \\
\cdot (\partial^2 \delta_{\mu\nu} - \partial_{\mu}\partial_{\nu})
A^j_{\nu}(x, \{a,a^*\})
\label{1}
\end{eqnarray}
where $px = \omega_p t - \vec{p} \vec{x}\;\;, \;\; S(\{a, a^*\})$ is
the so-called normal symbol of $S$ - matrix, the subscript "tree"
indicates the tree approximation, the subscript "c" means that only
connected part is included in the amplitudes, $a^j_{\lambda}(p)$
($a^{*j}_{\lambda} (p)$) stands for the  symbol of annihilation
(creation) operator of the gluon state with momentum $p$, color index
$j$ and a polarization $\lambda$,  $\varepsilon^{\lambda}_{\mu}$ is a
four-vector defining the polarization
$\lambda$ $(\varepsilon^{\lambda}_{\mu} p^{\mu} = 0)$ and $A^j_{\nu}$ is
a solution of the classical equation of motion. The $\{a , a^* \}$ -
dependence of $A^j_{\nu}$ comes via the Feynman-type boundary
asymptotic condition:
\begin{equation}
A^j_{\nu}(x , \{a, a^*\}) = A^{0j}_{\nu}(x, \{a, a^*\}) + O(g)
\label{2}
\end{equation}
where $g$ is a coupling constant and $A^0$ is a solution of the free
(without nonlinearity) equation of motion,
\begin{equation}
A^{0j}_{\nu} = \int \frac{d^3k}{(2\pi)^32\omega_k}
(a^j_{\lambda}(k)\varepsilon^{\lambda}_{\nu} e^{-ikx} +
{a^*}^j_{\lambda}(k) \bar{\varepsilon}^{\lambda}_{\nu} e^{ikx})\;.
\label{3}
\end{equation}
 Some comments are in order here:

 1) the condition (\ref{2}) fixes the solution of the classical
 equation of motion uniquely, provided the perturbation theory is
 well-defined, and due to this uniqueness a solution matching the
condition (\ref{3}) is the appropriate one;

 2) the power expansion of the formula (\ref{1}) in $a, a^*$ produces
 the multigluon amplitudes (with one punctured  gluon) in the obvious
 way;

 3) the $A^0$ in (\ref{3}) need not include all the $a$ and $a^*$ in
 the theory, in this case the formula (\ref{1}) generates a subclass
 of the tree amplitudes, corresponding to the  $a$,$a^*$  kept;

 4) the punctured particle in (\ref{1}) (corresponding to
 $a^j_{\lambda}(p)$) need not enter the $A^0$ in (\ref{3}), moreover,
 it need not even  be on-shell, in the latter case the formula
 (\ref{1}) generates rather formfactors than the amplitudes and is
 not gauge invariant;

 5) when the  punctured particle is taken on shell,  the amplitude is
 nonzero only if the solution $A$ in
 (\ref{1}) is sufficiently singular;
 
6) to have a possibility to keep two particles off shell or off the
 asymptotic condition (\ref{3}) one should  take the variation of
 (\ref{1}) with respect to another $a$, the $A$ in (\ref{1}) is then
 substituted by $\frac{\delta A}{\delta a}$ which obviously obeys the
 variation of the  classical equation of motion (at this point is
 essential that only connected  part of the amplitudes are considered).

 One can not proceed further in all generality as the YM
 equation is not integrable and one cannot find its solution for
an arbitrary boundary condition (\ref{3}). However, according to the comment 3),
 one can keep such a subset of $a$'s and $a^*$'s in (\ref{3})  that
 the corresponding  $A$ obeys simpler equation than the YM
 one. In the massive theory one proceeded keeping only space-uniform subset of
 $a$'s and $a^*$'s, the classical equation of motion reducing to an
 ordinary differential equation. That was the basic idea behind the
 threshold amplitudes activity mentioned in the introduction. Here
 I take $A^0$ such that $A$ obeys the SD equation. To be rigorous,
 note that in Minkovsky space (where we are) the $*$-operator squares
 to -1 and the SD equation says that the curvature  from is a
 $*$-eigenform with eigenvalue $i = \sqrt{-1}$, but it is not
 crucial, as after  the reduction to a subset of $a$'s and $a^*$'s in
 (\ref{3}) the $A^0$ and $A$ need not be real.  The appropriate
 solutions can actually be obtained by the Wick's analytical
 continuation. Using the (analytically continued) 't Hooft anzatz
 \begin{equation}
 A = \frac{i}{g} \bar{\Sigma}_{\mu\nu} \partial_{\nu} \ln \Phi
 dx^{\mu}
 \label{4}
 \end{equation}
 the SD equation is known to reduce to the following equation on
 $\Phi$:
 \begin{equation}
 \frac{1}{\Phi} \Box \Phi = 0
 \label{5}
 \end{equation}
 ( $\partial_{\nu}$ and $ dx^{\mu}$ include   $i = \sqrt{-1}$ where it is necessary
according to the Wick's rotation rules, $\bar{\Sigma}_{\mu\nu}$ are the 't Hooft matrixes,
 $\bar{\Sigma}_{\mu\nu} = - \bar{\Sigma}_{\nu\mu}\;, \;
 \bar{\Sigma}_{a 0} = -\frac{\sigma^a}{2}\;, \bar{\Sigma}_{ab} =
 \varepsilon^{iab} \frac{\sigma^i}{2}\;,  \; \sigma^i$ are the Pauli
 matrixes).
 For the current purposes I take the solution
 \begin{equation}
 \Phi = 1 + g\int \frac{ d^3 q}{(2\pi)^3 2q_0} ( \alpha(q)e^{-iqx} +
 \alpha^*(q)e^{iqx})
 \label{6}
 \end{equation}
 with $ q^2_0 - (\vec{q})^2 = 0$.
 Substituting $\Phi$ into (\ref{4}) and expanding in $g$ up to first
 order relates $\alpha, \alpha^*$ to $a, a^*$. At this point I need
 to describe special gluon states arising in the asymptotics of the
 solution (\ref{6}), (\ref{4}). This will define the subclass of the
 multigluon amplitudes for which this solution plays the role of
 generating function.

  Consider right triple of orthogonal 3-vectors, $(\vec{n}_1, \vec{n}_2,
  \vec{q}$), $\vec{q}$ being the spatial part of $q$ in (\ref{6}).
  Introduce circular polarization ${\varepsilon}^-(n, q) =(0, \vec{n}_1
  - i\vec{n}_2)$. Introduce also matrixes  $\sigma_{n_{i}}(nq) =
  \sigma^kn_{i}^{k}$ and  a color orientation $\sigma^+(n,q) =
  \sigma_{n_1} + i\sigma_{n_2}$.
  That are these gluon states, with the circular polarization
  $\vec{\varepsilon}^-(n,q)$ and the color polarization $\sigma^+(n,q)$
  which will appear in the asymptotic states. $a_{+,-}(q)$ and
  $a^*_{+,-} (q)$ will stand for the corresponding annihilation and
  creation symbols (notice that the states are independent of the choice of
   $(\vec{n}_1, \vec{n}_2$ at given $q$ and the time  direction. The parameters
  $\alpha, \alpha^*$ from (\ref{6}) read then as follows:
  \begin{eqnarray}
  \alpha(q) = \frac{2 a_{+,-} (q)}{iq_o} \nonumber \\
  \alpha^*(q) = \frac{2ia^*_{+,-}(q)}{q_0}
  \label{7}
  \end{eqnarray}

The transversal part of the solution $A$ (which only contributes in
  (\ref{1})) is
  \begin{eqnarray}
  A^{T}_k &=& \frac{1}{\Phi} \int \frac{d^3q}{(2\pi)^3 2q_0}
  [a_{+,-}(q) \sigma^+(n,q) \varepsilon^-_k(u,q) e^{-iqx} + \nonumber \\
  &+& a^*_{+,-}(q) \sigma^+(n,q) \varepsilon^-_k(n,q) e^{iqx}]
  \label{8}
  \end{eqnarray}
  One immediately sees that the solution (\ref{8}) does not have any
  singularity which could give a nonzero amplitude when the punctured
  gluon is on-shell (independently of its polarization and color
  orientation). This is the nullification cited above. When the
  punctured gluon is off-shell, the formula  (\ref{8})
  substituted into (\ref{1}) generates nonzero formfactors (which are not gauge invariant, of course).
  As was said in comment  5), to find the amplitudes with two
  arbitrary polarized and color oriented gluon and arbitrary number
  of the specially polarized and oriented gluons one needs to solve
  the variation of the YM equation in the background of $A(x, \{a,
  a^*\})$. This will be done  elsewhere.

  3. I would like to indicate what developments are possible on the basis
  of exposed in this note.

  1) SU(N) case allows similar construction. In that  general case
  the spinor helicity formalism comes naturally via the twistor
  description of solutions of SD equation.

  2)The variation of the YM equation can be
  solved in the background of $A(x, \{a, a^*\}$ (even in SU(N) case)
  which will allow to construct the generating function for the
  processes with two arbitrarily polarized gluons and arbitrary
  number of the specially polarized ones. Actually, the recursive
  relations used in  \cite{PT}, \cite{BG} to find this type of
  amplitudes must be just a perturbative expansion of the variation of YM.

  3) Constructing the Green function in the background of $A(x, \{a
  a^*\}$ will allow to develope a perturbation theory for any
  processes beyond the tree approximation including arbitrary number of special gluons, similar to
  \cite{smith}, \cite{arg} in the threshold amplitudes activity.
  I would like to recall that using this kind of technique the
  authors of \cite{bez} (see also \cite{GV}) managed to collect all
  loops contributions in the leading term of the threshold amplitude
  when the number of particle is of the order of inverse coupling
  constant.

  4) The nullification of the amplitudes with the specially polarized
  gluons was explained  by an effective supersymmetry 
  Ward identities \cite{Su} (the idea was that
  the normal YM can be completed up to the supersymmetric one, while the
  superpartners do not contribute at the tree level). A similar
  interpretation of the nullification in the threshold amplitudes
  could easily  be possible.

  5) In \cite{S} a general approach based on the Whitham technique
  was developed to treat arbitrary next-to-threshold amplitudes. By the way,
  the effective theory was nicely formulated in terms of the modular
  geometry and happened to be solvable at the classical level. I
  hope that an analogous construction is possible in the present
  case. The corresponding effective theory would describe processes
  including the specially polarized gluons witih arbitrary momenta
  and arbitrary polarized soft gluons.

It's my pleasure to thank I.Kogan who informed me about the multigluon amplitudes
activity. This work was partially supported by INTAS grant 93-633.

  \end{document}